\documentstyle[12pt]{article}
\newcommand{\doublespace}{\renewcommand{\baselinestretch}{1.75}
   \Large\normalsize}

\renewcommand{\ref}[1]{\raisebox{.6ex}{[#1]}}

\newcommand{\be}{\begin{equation}}
\newcommand{\ee}{\end{equation}}

\begin{document}

\doublespace

\title{Dynamical Correlation Theory for an Escape Process }

\author{Ping Ao                               \\
Department of Physics, FM-15                  \\
University of Washington, Seattle, WA 98195, USA  \\ }

\maketitle

\begin{abstract}
A dynamical theory which incorporates the electron-electron correlations
and the effects of external magnetic fields for an electron escaping
from a helium surface is presented.
The degrees of freedom in the calculation of the escape rate is reduced from
$3N$ to 3 as compared with other approach.
Explicit expressions for the escape rate in various situations
are obtained. In particular, in the weak parallel magnetic field limit the
tunneling rate has an exponential dependence quadratic with magnetic field
strength and an unusual exponential increase linear with temperature.
\end{abstract}

\noindent
PACS${\#}$s: 73.40.Gk; 71.45.Gm; 73.20.Dx

\newpage

When many electrons sit in a metastable well near a helium surface,
the escape of an electron
from the well is no longer a single particle problem
due to the Coulomb interaction. Electron-electron (e-e) correlations
play an essential role in the escape process
and the physics becomes very rich.
This system is ideal to test our
understandings of the escape from a metastable well when many body effects are
important.
A number of experiments have been performed on the escaping of electrons from a
helium surface.\ref{1-5}
Recent theoretical studies have mainly concentrated
on how the static e-e correlations affect the escape rate.
A comprehensive understanding has been obtained
in treating the e-e correlations
as either instantaneous following the motion of an escaping electron
or not following it at all.\ref{6-11}
The real dynamical
nature of the e-e correlations, however, has not been explored adequately.
As one can see from a rather similar problem of the escaping of a particle
from a metastable well in the presence of an environment,
the dynamical response can be dominant.\ref{12,13}
In the present paper we develop a dynamical theory for the escaping of an
electron from a helium surface to account for the effects of
both static and dynamical e-e correlations and magnetic fields.
The original $3N$ degree freedom problem\ref{10}
is simplified to a 3 degree freedom
one, with $N$ the total number of electrons in the problem. Expressions
in various limits for the escape rate in the terms of temperature, 2-d
electron density, and external magnetic fields are obtained.
Particularly, an unusual exponential increase linear with temperature
of the tunneling rate in the presence of a weak parallel magnetic field
is found. The dependence on the magnetic field is quadratic in exponent
in this limit.

We consider the experimental relevant situation in which
the lifetime of the metastable state of an electron is much longer
than the relaxation time of the 2-d electrons
and the density of 2-d electrons is low such that the Fermi temperature
is the smallest energy in the problem. The escaping events are then
statistically independent of each other and the exchange effect
of an escaping electron with 2-d electrons can be ignored.
A separation between the escaping electron and the remaining 2-d electrons
for each escape event can be made.
For simplicity, we shall ignore the weaker interactions of the escaping
electron to the surface waves of liquid helium and the helium vapor
atoms, which has been discussed elsewhere\ref{11}.

We start with the derivation of the effective Hamiltonian for the escaping
electron classically.
This Hamiltonian will be the base to study the escape process.
For quantum tunneling, an imaginary time path integral method following
Ref.[13] will be used.
The classical equation of motion for the escaping electron is:
\be
   m\frac{d^{2}}{dt^{2}} {\bf R}_{t} = - \nabla V_{0}(z_{t})
      + \frac{e}{c} \frac{d {\bf R}_{t}}{dt} \times {\bf B}_{ex}
      + e  {\bf E}_{in}({\bf R}_{t}) \; ,
\ee
with $ V_{0}(z) = V_{w}(z) + V_{i}(z) + V_{n}(z)$.
Here $m$ is the mass of an electron, $e$ is the negative electron charge,
$c$ is speed of light,
and $V_{w}$ is the hardwall potential, $V_{w} = \infty$
for $z<0$,  $V_{w} = 0$ for $z>0$. This potential mimics the fact
that it costs energy $\sim 1$eV for an electron to go into the liquid helium,
which is very large in the present problem.
The image potential $V_{i}$ due to the polarization of helium liquid is
$ V_{i}(z) = - e^{2}\Lambda /z $,
with $\Lambda = (\epsilon - 1 )/4(\epsilon + 1 ) $. The dielectric constant of
liquid helium $\epsilon = 1.057$.
The potential $V_{n}$ is the total electric potential
produced by the external applied electric field (perpendicular only) and the
electric field produced by the mean density  $n_{0}$ of the 2-d electrons,
\be
   V_{n}(z) = - e[E_{ex} + 2\pi e (1-4\Lambda) n_{0}]z \; .
\ee
The condition for 2-d electron escaping to $z=\infty$ from the surface is
$E_{ex} + 2\pi e (1-4\Lambda)n_{0} < 0$. The external applied
magnetic field is ${\bf B}_{ex}$, and ${\bf E}_{in}$ is the
electric field produced by the 2-d electron density
deviated from $n_{0}$.

The induced electric field ${\bf E}_{in}$ is generated by the 2-d
density deviation from the mean value $n_{0}$,
which in turn is induced by the escaping electron.
Consequently, ${\bf E}_{in}$ can be expressed
in terms of the motion of the escaping electron. The procedure is as follows.
Let the 2-d electron fluid sit on the surface of the liquid helium,
the $x-y$ plane. The fluid is described by a set of hydrodynamical equations:
the continuity equation and the Euler's equation.
In the small density deviation
and the nonrelativistic limit, we linearize the hydrodynamical equations.
We solve for the density deviation, which is presented by the motion of the
plasma modes. Then using
the Poisson equation we obtain the induced electric field as\ref{11}
\[
   {\bf E}_{in}({\bf R}_{t}) =  - \nabla \int d{\bf k} \int d\omega \;
             \frac{n_{0}e^{3} }{m}
             \frac{1}{\omega^{2} - \omega_{P}^{2}(k) }
             \int \frac{dt'}{2\pi}
             \exp\{-kz_{t'}\}
             \exp\{-i({\bf k}\cdot{\bf r}_{t'} - \omega t')\}
\]
\be
             (1-4\Lambda )^{2} \exp\{-kz_{t}\}
             \exp\{i({\bf k}\cdot{\bf r}_{t} - \omega t) \} .
\ee
Here the plasma dispersion relation $\omega_{P}(k)$ in eq.(3) is
\be
   \omega_{P}^{2}(k) = \omega_{B}^{2} + \frac{n_{0}e^{2} }{m} 2\pi k
                      (1-4\Lambda ) + \frac{k_{B}T }{m} k^{2} ,
\ee
and the cyclotron frequency $\omega_{B} = eB_{\bot ex} /mc $ with
$B_{\bot ex}$ the component of the external magnetic field perpendicular
to the helium surface.
In the calculation the pressure $p = n k_{B}T$
for the 2-d classical electron fluid phase has been used.

The induced electric field is contributed by the response of the
environment, the plasma modes of 2-d electrons,
to the motion of the escaping electron.
Hence we are dealing a problem similar to the one in the discussion of the
macroscopic quantum effect\ref{13}, where the total Hamiltonian has three
parts, a dissipative environment consisting of harmonic oscillators,
a system of interesting, and the coupling between the
system and the environment.
Using this analogy, we find that the following effective
Hamiltonian is equivalent to eqs.(1,3) to describe
the motion of the escaping electron:
\[
  H = \frac{1}{2m}\left[ {\bf P} - \frac{e}{c} {\bf A}_{ex} \right]^{2}
       + V_{A}(z)
       + \int_{k < g\sqrt{n_{0}} } d{\bf k} \sum_{j=1,2}
       \left[ \frac{1}{2m} p_{j}^{2}({\bf k})
       + \frac{1}{2} m \omega_{P}^{2}(k) \times  \right.
\]
\be
       \left.    \left( q_{j}({\bf k})
       - \frac{e \sqrt{ n_{0} e^{2} } (1-4\Lambda )\; \exp\{-kz\} c_{j} }
              {m\omega_{P}^{2}(k) } \right)^{2}  \right]  ,
\ee
with
\be
    V_{A}(z) = V_{0}(z) - \frac{1}{2} \int_{0}^{g\sqrt{n_{0}} } dk \; 2\pi k
       \frac{n_{0} e^{4} (1-4\Lambda )^{2} }{m\omega_{P}^{2}(k) } \;
       \exp\{-2kz\} \; ,
\ee
and $c_{1}=\cos({\bf k}\cdot{\bf r}) $, $c_{2}=\sin({\bf k}\cdot{\bf r}) $,
${\bf B}_{ex} = \nabla\times{\bf A}_{ex} $.
Here $g$ is a numerical factor of order unit to be discussed below.
The choice of the effective Hamiltonian is chosen in a form as in
Ref.[14] in the discussion of the dissipative bath in quantum tunneling.
When 2-d electrons follow the motion of the escaping electron completely,
the so called the adiabatic limit, the response of 2-d electrons
is described by the adiabatic potential given by eq.(6).
The deviation from the adiabatic response is described by the plasma dynamics,
the last term in eq.(5).
The second term of eq.(6) corresponding to the correlation potential discussed
in Ref.[1]. Hence an alternative justification of its usage in Refs.[1-5]
is obtained here.
Several features of the effective Hamiltonian should be pointed out.
The coupling between the escaping electron and the plasma is highly nonlinear
in the coordinate of the escaping electron. The damping of the escaping
electron due to plasma is clearly superohmic case,
when $B_{\bot ex} \neq 0$.\ref{15,13,11}
It is also superohmic when $B_{\bot ex} = 0$, a situation similar to the
polaron problem.
The effective Hamiltonian contains a weak temperature dependent
through the plasma
frequency, which comes from the equilibrium state of the electron fluid.
The potential $V_{A}(z)$ is influenced by $B_{\bot ex}$
through the plasma frequency $\omega_{B}$ dependence.

Because the average distance between  electrons is $1/\sqrt{n_{0}}$,
there is no plasma mode of large wavenumbers $k >> \sqrt{n_{0}}$.
Furthermore, the 2-d electron density deviation is large in this regime
and the linearization approximation leading to eqs.(3,4) is not accurate.
Then we need to introduce
a cutoff plasma frequency $\omega_{c} = \sqrt{2\pi g n_{0}^{3/2} e^{2}/m }$
corresponding to $k \sim g \sqrt{n_{0}}$
which determines the fastest response of the plasma, as shown in Ref.[16].
This would suggest that we could only determine the numerical
factor $g$ in eq.(6) to be an order of unit, and
the radius of the hole created by the escaping electron pushing 2-d electrons
sideward\ref{1} to be the order of $1/\sqrt{n_{0}}$.
However, $g$ can be theoretically determined accurately in the following way:
Let the escaping electron sit in the center of the hole created by itself in
2-d electrons. Because of the rotational
symmetry there is no net force from 2-d
electrons acting on the escaping electron. The escaping electron only feels
the externally applied force, {\it i.e.},
$-\nabla V_{A}|_{hole \; center } = e E_{ex} \hat{z}$. The choice of $g$ must
satisfy this condition. Therefore we have $g = 2\sqrt{\pi} $.
This completes the derivation of the effective Hamiltonian eq.(5).

The effective Hamiltonian may have a wider application region than
the region of validity of the hydrodynamical approach,
so long as elementary excitations,
such as the plasma modes calculated above, dominate the response
of the 2-d electrons to the escaping electron.
For example, this may include the case of the Wigner lattice phase.
In this case, the plasma modes will be replaced by the phonon modes.
One can expect that the form of the effective hamiltonian
is the same as eq.(5) in the long wave length limit when $B_{\bot ex}=0$,
because the density deviation is the same longitudinal one.
As the local structure of 2-d electrons is not
changed in the Kosterlitz-Thouless transition, we then expect the same
short distance behavior. Then there should be no change of the adiabatic
potential therefore no change of the escape rate cross
the melting temperature. This is consistent with a recent experiment.\ref{4}

We now calculate the escape rate starting from eq.(5) for various situations.
In the high temperature regime, the escape is dominated by the thermal
activation. The escape rate is
$ \Gamma_{T} = A \; \exp\{-{\cal E}_{b}/k_{B}T \} $.
Here the prefactor $A$ is weakly temperature dependent
and is proportional to the density of 2-d electrons.
The barrier height ${\cal E}_{b}$ is determined by the equation
$ {\cal E}_{b} = V_{A}(z_{max}) - {\cal E}_{0} $,
where $z_{max}$ is the position of the barrier top.
The metastable (ground) state energy ${\cal E}_{0}$ is directly calculated
from the effective Hamiltonian of eq.(5) as
\be
   {\cal E}_{0} = - \frac{e^{2}\Lambda }{2a_{B}' }
                  - \frac{1}{2}\int_{0}^{g\sqrt{n_{0}} }dk
                    \frac{2\pi k e^{4}n_{0} }{ m\omega_{P}^{2}(k) }
                    (1-4\Lambda)^{2} \; ,
\ee
with $a_{B}'= \hbar^{2}/me^{2}\Lambda $.
Much weaker influences from the Stark shift and vertical spread of 2-d
electrons have been ignored.
If we ignore its small effect on the ground state energy,
there is no influence of the parallel magnetic field $B_{\| ex}$
on the thermal activation rate.
On the contrast, the activation rate will be increased in the presence of a
perpendicular magnetic field $B_{\bot ex}$,
because $B_{\bot ex}$ affects the potential $V_{A}(z)$ in a special way.
Specifically, we find that the barrier height ${\cal E}_{b}$ in the weak field
limit as:
\be
   {\cal E}_{b} =
    \left\{  \begin{array}{ll}
        {\cal E}_{b}|_{ B_{\bot ex}=0 }
        - 2\sqrt{\pi n_{0} } \; z_{max} \left(\frac{\omega_{B} }{\omega_{c} }
        \right)^{2}
        (1-4\Lambda) 2 e^{2}  \sqrt{\pi n_{0} }  \; , &
        2\sqrt{\pi n_{0} } \; z_{max} << 1  \; ,                       \\
        {\cal E}_{b}|_{ B_{\bot ex}=0 }
        - \left(\frac{\omega_{B} }{\omega_{c} }\right)^{2}
        \ln\left(\frac{\omega_{c} }{\omega_{B} }\right){\ }(1-4\Lambda) 2 e^{2}
        \sqrt{\pi n_{0} }  \; ,&
        2\sqrt{\pi n_{0} } \; z_{max} >>1  \; .
    \end{array} \right.
\ee
This result may be interpreted as that $B_{\bot ex}$
effectively reduces the interaction between the 2-d electrons
and the escaping electron.
For a strong magnetic field we should pointed out that
the short magnetic length scale introduced by the strong magnetic field
may indicate the invalid of the present hydrodynamics approach.

In the low temperature region quantum tunneling dominates the escape.
The tunneling rate is
$   \Gamma_{Q} = \omega_{a}\; \exp\{-S_{c}/\hbar\}$.
The prefactor $\omega_{a}$  is of an order of ${\cal E}_{0} /\hbar $.
The classical action $S_{c} = S_{eff}[{\bf R}_{c}(\tau)]$ is evaluated at
the classical trajectory ${\bf R}_{c}(\tau)$ determined by the equation
$ \delta S_{eff}[{\bf R}(\tau)] = 0 $ which is a 3 instead of $2N+1$
dimensional
partial differential equation, with the effective action
$S_{eff}$ as\ref{13}
\[
   S_{eff}[{\bf R}(\tau)] = S_{0}[{\bf R}(\tau)]
           + \frac{1}{2} \int d\tau \int d\tau' \int d{\bf k}
             \frac{n_{0}e^{4}(1-4\Lambda)^{2} }{4m\omega_{P}(k) } \;
           \exp\{ - \omega_{P}(k)|\tau - \tau'| \}
\]
\be
      \left\{ \left[ \exp\{-kz(\tau)\} - \exp\{-kz(\tau')\} \right]^{2}
      + 2\exp\{-k(z(\tau)+z(\tau'))\}
        [1- \cos[{\bf k}\cdot({\bf r}(\tau) - {\bf r}(\tau') ) ]] \right\} ,
\ee
with
\be
   S_{0}[{\bf R}(\tau)] = \int d\tau \left[ \frac{1}{2} m\dot{\bf R}^{2}(\tau)
           + i \frac{e}{c} {\bf A}_{ex}\cdot \dot{\bf R}
           + V_{A}(z(\tau)) \right] .
\ee
The nonlocal term in time in eq.(9)
is a result of the reduction of the degrees of freedom from $3N$ to 3.
Numerical calculation can be used in order to make a detailed comparison to
experiments. In the following we discuss some prominent features of the
tunneling rate analytically.

First, we consider the case of the zero parallel magnetic field $B_{\| ex}=0$.
If we drop the last and non-negative term in eq.(9),
replace $\dot{\bf R}^{2}$ by $\dot{z}^{2}$ in eq.(10), and apply
the 1-d WKB approximation, we obtain
a lower bound of the classical action $S_{c}$:
\be
   S_{lower} = 2 \int_{z_{1}}^{z_{2}} dz
             \sqrt{ 2m[V_{A}(z) - {\cal E}_{0} ] } \; ,
\ee
with $z_{1}$ and $z_{2}$ the turning points which are the solutions of the
equation $V_{A}(z) - {\cal E}_{0} = 0 $.
The tunneling rate calculated in this
way gives the upper bound for the tunneling rate as pointed out in Refs.[10,11]
This upper bound of tunneling rate corresponds to the physical situation in
which the 2-d electrons follow the motion of the escaping electron
instantaneously, the adiabatic limit, as a general theorem has
shown.\ref{13,17}
The classical trajectory corresponding the case in which
${\bf r}(\tau) = constant$ therefore
$1-\cos[{\bf k}\cdot({\bf r}(\tau)-{\bf r}(\tau')]=0$ ( note $B_{\| ex}=0$).
Setting $ \exp\{-kz(\tau') \} =  1$, integrating over $\tau'$,
and again using the 1-d WKB approximation,
we obtain an upper bound for $S_{c}$ from eq.(9):
\be
   S_{upper} = 2 \int_{z'_{1}}^{z'_{2}} dz \sqrt{ 2m[V_{U}(z)-{\cal E}_{0}]}\;
,
\ee
with $z'_{1}$ and $z'_{2}$ the turning points which are the solution of the
equation $ V_{U}(z) - {\cal E}_{0} = 0 $. Here the potential $V_{U}(z)$ is
\be
   V_{U}(z) = V_{A}(z) + \frac{1}{4} \int_{0}^{g\sqrt{n_{o} } } dk
    \frac{2\pi k e^{4} n_{0} (1-\Lambda)^{2} }{m\omega_{P}^{2}(k) }
    ( \exp\{ -kz \} - 1)^{2} \; .
\ee
This upper bound for $S_{c}$ is smaller than that
given by the frozen potential\ref{10,11}, and is a better one.
This is due to the partial inclusion of the dynamics of the plasma modes.

The tunneling rate increases as $ B_{\bot ex}^{2} $ in the low field limit
because of the suppression of the correlation potential
discussed in the case of thermal activation rate.
There is another reason for the increase of tunneling rate, the increase of
the adiabaticity:
If we fix the adiabatic potential $V_{A}(z)$, and allow
the plasma frequencies to increase independently,
then the e-e response will be able to follow
the motion of escaping electron more closely,
and $S_{c}$ decreases towards its lower bound,
$S_{c} \rightarrow S_{lower}$, the adiabatic limit.
This result is the opposite of Ref.[10].
Because the damping is superohmic, the temperature dependence
due to damping in the tunneling regime is found to behave as $O (T^{7}) $
and is small in the low temperature limit.\ref{11,15}

Now we discuss the case that the parallel magnetic field is present,
$B_{\| ex} \neq 0$ and $B_{\bot ex}=0$.
By choosing the parallel
magnetic field along the $x$-direction and
${\bf A}_{ex} = (0, -B_{\| ex}z, 0)$,
according to the effective Hamiltonian eq.(5) we have the first two terms as
\be
   H_{0} = \frac{1}{2m} P_{z}^{2} + V_{A}(z)
      + \frac{1}{2m} \left[ P_{y} + \frac{e}{c} B_{\| ex} z \right]^{2} \; .
\ee
The $x$-direction motion is irrelevant. This Hamiltonian describes the 1-d
motion of a particle coupling to a harmonic oscillator
with a zero frequency.
We integrate over the $y$ coordinate to obtain the
effective action for the motion in $z$ direction as
\be
   S_{eff}[z(\tau)] = \int d\tau \left[ \frac{1}{2} m\dot{z}^{2}(\tau)
                      + V_{A}(z(\tau)) \right]
   + \frac{1}{4m} \left(\frac{eB_{\| ex} }{c} \right)^{2} \frac{k_{B}T}{\hbar}
          \int d\tau \int d\tau' \; [ z(\tau) - z(\tau') ]^{2} \; .
\ee
This result suggests that the $y$ direction motion acting as a dissipative
environment to the $z$ direction motion.\ref{18}
The influence action from the plasma modes is ignored by assuming the
adiabatic limit for simplicity, which will be discussed below.
In the small field limit, the classical trajectory remains unchanged.
The classical action can then be evaluated perturbatively:
\be
   S_{c} = S_{lower} + \frac{1}{2m} \left(\frac{eB_{\| ex} }{c} \right)^{2}
              \int d\tau \;  z_{c}^{2}(\tau) -
   \frac{1}{2m} \left(\frac{eB_{\| ex} }{c} \right)^{2} \frac{k_{B}T}{\hbar}
           \left(\int d\tau \;  z_{c}(\tau) \right)^{2} \; ,
\ee
with semiclassical trajectory $z_{c}(\tau)$ determined by the usual equation
$\frac{1}{2} m \dot{z}_{c}^{2} = V_{A}(z_{c}) - {\cal E}_{0}$.
The tunneling rate decreases at zero temperature because of the bending of
classical trajectory, and increases
exponentially with temperature because of the excitation of the $y$-direction
motion, with a $B^{2}_{\| ex}$ dependence on the parallel magnetic field.
In the calculation we have ignored a weak dependence of ${\cal E}_{0}$ on
$B_{\| ex}$.

We have discussed the crossing from thermal activation to quantum tunneling,
the effect of the correlation potential, the dynamical response of the
2-d electrons to the escaping electron, and the effect of weak parallel
magnetic field. We now discuss the conditions to observe these effects.
Let $\omega_{b}^{2} = |V_{A}''(z_{max})|/m$ be the small
oscillation frequency in the inverse potential $ - V_{A}(z)$.
The crossing from thermal activation to quantum tunneling occurs at
$k_{B}T_{0} = \hbar\omega_{b}$ according to the standard theory.\ref{15}.
The frequency $\omega_{b}$ also determines the dynamics of the escaping
electron in the inverse potential.
In order for the 2-d electrons to follow the motion of the escaping electron,
the cutoff plasma frequency should satisfies the inequality
$\omega_{c} > \omega_{b}$, and then the tunneling is
essentially an adiabatic process\ref{11,13,14}. Similarly,
the condition for the weak parallel magnetic
field $B_{\| ex}$ is  $eB_{\| ex}/mc < \omega_{b}$.
To manifest the correlation effect in the escape process,
we need $\sqrt{n_{0}} z_{max} \geq 1$ in the case of the thermal
activation, and $\sqrt{n_{0}} z_{2} \geq 1$ in the case of quantum tunneling,
otherwise the correlation effect is small.

Finally, let us make a brief comparison to experiments.
The present theory is in agreement with experiments in the thermal activation
regime, where consistent experimental data exist.\ref{11}
The existing quantum tunneling data in the absence of magnetic field\ref{3,4}
do not agree with each other and there is no satisfactory explanation.
However,
the magnetic field dependent tunneling results obtained in the present paper,
the quadratic dependence on the parallel magnetic field and linear in
temperature, agree with a recent experiment described in Ref.[5].

{\bf Acknowledgements}:
The author thanks A.J. Leggett for suggesting this
problem and for sharing of his physics insight.
Discussions with A.J. Dahm, J.M. Goodkind, J. Rammer,
and at various stages of the work are appreciated.
The author is particularly grateful to E.Y. Andrei for bringing the data
of Ref.[5] to his attention prior to its publication.
This work was supported by
the John D. and Catherine T. MacArthur Foundation at University of Illinois,
and by the US National Science Foundation under Grant \# DMR 89-16052.

\noindent
{\bf References}

\noindent
1. Y. Iye, K. Kono, K. Kajita, and W. Sasaki,
    J. Low Temp. Phys. {\bf 38}, 293 (1980).  \\
2. K. Kono, K. Kajita, S.I. Kobayashi, and W. Sasaki,
      J. Low Temp. Phys. {\bf 46}, 195 (1982).  \\
3. J.M. Goodkind, G.F. Saville, A. Ruckentein, and P.M. Platzman, Phys.
      Rev. {\bf B 38}, 8778 (1988).  \\
4. E.Y. Andrei, S. Yucel, and L. Menna, Phys. Rev. Lett. {\bf 67}, 3704
     (1991). \\
5. S. Yucel, L. Menna, and E.Y. Andrei, J. Low Temp. Phys. {\bf 89}, 257
     (1992); L. Menna, S. Yucel, and E.Y. Andrei, preprint.   \\
6. S. Nagano, S. Ichimaru, H. Totsuji, and N. Itoh, Phys. Rev. {\bf B 19},
     2449(1979).   \\
7. Y.M. Vilk and Y.P. Monarhkar, Fiz. Nizk. Temp. {\bf 13}, 684 (1987)
     [Sov. J. Low Temp. Phys. {\bf 13}, 392 (1987)].  \\
8. M.Y. Azbel, Phys. Rev. Lett. {\bf 64}, 1553(1990).                \\
9. S. Y\"{u}cel and E.Y. Andrei, Phys. Rev. {\bf B42}, 2088 (1990).   \\
10. M. Azbel and P.M. Platzman, Phys. Rev. Lett. {\bf 65}, 1376(1990).  \\
11. P. Ao, Ph.D. thesis, chapter 3, University of Illinois at Urbana-Champaign,
      1991 (unpublished).   \\
12. H.A. Kramers, Physica {\bf 7}, 284 (1940).   \\
13. A.O. Caldeira and A.J. Leggett, Ann. Phys. (N.Y.) {\bf 149}, 374 (1983);
      {\bf 153}, 445(1984)(E).  \\
14. A.J. Leggett, Phys. Rev. {\bf B 30}, 1208 (1984).     \\
15. For a recent review, see for example,
   P. H\"{a}nggi, P. Talkner, and M. Borkovec,
   Rev. Mod. Phys. {\bf 62}, 251(1990).              \\
16. K.I. Golden, G. Kalman, and P. Wyns, Phys. Rev. {\bf A41}, 6940 (1990).
\\
17. A.J. Leggett and G. Baym, Phys. Rev. Lett. {\bf 63}, 191 (1989).   \\
18. A.J. Leggett, in {\it Granular Nanoelectronics}, edited by D.K. Ferry,
J.R. Barker, and C. Jacobini, Plenum, New York, 1991.

\end{document}